\documentclass[12pt]{article}

\usepackage{fullpage, hyperref,bbm,amssymb,amsfonts,amsthm}
\usepackage[noend]{algorithmic}
\usepackage{algorithm}
\usepackage{amsmath}
\usepackage{setspace}
\usepackage{cleveref}
\usepackage{color}
\usepackage{latexsym}
\usepackage{verbatim} 
\usepackage{graphics}
\usepackage{graphicx}
\usepackage{tikz}
\usepackage{multicol}
\usetikzlibrary{automata,positioning}

\newcommand{\kets}[1]{| #1 \rangle}                 
\newcommand{\bras}[1]{\langle #1 |}                 
\newcommand{\braket}[2]{\langle #1 | #2 \rangle}         

\newtheorem{theorem}{Theorem}
\newtheorem{lemma}{Lemma}

\begin{document}
\title{Optimizing Quantum Walk Search on a Reduced Uniform Complete Multi-Partite Graph} 
\author{Chen-Fu Chiang
		\thanks{Department of Computer Science, State University of New York Polytechnic Institute, Utica, NY~13502, USA.
		Email:\texttt{chiangc@sunyit.edu}},
		Chang-Yu Hsieh
		\thanks{Department of Chemistry, Massachusetts Institute of Technology, Cambridge, MA~02139, USA.
		Singapore-MIT Alliance for Research and Technology (SMART) Centre, Singapore 138602.
		Email:\texttt{changyuh@mit.edu}}
		}
		
\maketitle

\begin{abstract}		%
In a recent work by Novo et al. (Sci. Rep. $\mathbf 5$, 13304, 2015),
the invariant subspace method was applied to
the study of continuous-time quantum walk (CTQW). The method helps to reduce a graph into a simpler version that 
allows more transparent analyses of the quantum walk model. In this work, we adopt
the aforementioned method to investigate the optimality of a quantum walk search of a marked element on a 
uniform complete multi-partite graph. We formulate the eigenbasis that would facilitate the transport between the two lowest 
energy eigenstates and demonstrate how to set the appropriate coupling factor to preserve the optimality.
\end{abstract}

\section{Introduction}	   %
Various quantum computational frameworks, such as Quantum Circuit Model \cite{nielsen2002quantum}, 
Topological Quantum Computation \cite{freedman2003topological},  Adiabatic Quantum Computation (AQC) \cite{farhi2000quantum}, 
Quantum Walk (QW) \cite{aharonov1993quantum,farhi1998quantum}, Resonant Transition Based Quantum Computation (RTBQC) 
\cite{chiang2017resonant} and Measurement Based Quantum Computation (MBQC) \cite{briegel2009measurement} 
have been proposed to attack problems that are considered extremely difficult for classical computers. Notable
successes include the inventions of Shor's factoring algorithm and Grover's search algorithm, which manifest
indisputable enhancement over all known classical algorithms designed for the same purpose.
Among the proposed quantum computational frameworks above, quantum walk models are certainly among the most
heavily supported. They provide a natural framework for tackling spatial search problems such as implementing the 
Grover's search algorithm \cite{grover1996fast}. In addition, they are central to quantum algorithms \cite{childs2004spatial, childs2003exponential} 
created to tackle other computationally hard problems, such as graph isomorphism\cite{berry2011two, gamble2010two, douglas2008classical}, 
network analysis and navigation \cite{berry2010quantum, sanchez2012quantum}, and quantum 
simulation \cite{lloyd1996universal, berry2009black, schreiber20122d}, even including certain aspects of 
complex biological processes \cite{engel2007evidence, rebentrost2009environment}.  Furthermore, due to the simple physics
principle behind quantum walk models, various efforts have been made to establish a better understanding
of quantum walk models by relating to other major quantum computational frameworks or explore novel approaches
to exploit quantum walks to perform a greater variety of tasks \cite{childs2009universal, du2003experimental, wong2016irreconcilable, sanders2017qwchimera, kitagawa_12, zanetti_14}.\\

Quantum walks can be formulated in both discrete time \cite{aharonov1993quantum} and continuous time \cite{farhi1998quantum} versions. In 
this work, we focus on the study of continuous-time quantum walk (CTQW), not only because it offers a simpler physical picture but also it is 
less challenging to perform CTQW experiments in comparison to their discrete-time
counterparts.  Furthermore, if implementing CTQW in a quantum circuit model, robust quantum computations could
be attained due to the availability of fault tolerance and error corrections.
Based on these motivations, we set out to investigate how to optimize CTQW searches on a uniform complete multi-partite graph. 
Although uniform complete multi-partite graphs constitute just a subset of all possible graphs, they include some of the most important 
examples, such as complete graphs, complete bipartite graphs and star graphs which will be further elaborated in section \ref{sect:Examples}, 
in applications of quantum walks to computations. \\

In this work, we adopt the invariant subspace method from Ref.\cite{novo2015systematic}, which allows us to perform
a dimensionality reduction to simplify the analyses of CTQW on a uniform complete multi-partite graph. 
In short, the key is to transform the original graph to a much simpler structure yet retain pertinent properties that 
we would like to investigate, such as the optimality of a quantum walk search. In this way, the analysis becomes more 
transparent and the dynamics of the walker can be more intuitively understood on an abstract level. Throughout the text, 
we also refer to a multi-partite graph as a $P$-partite with a slight twist on the standard notation. The difference is that the 
whole graph has actually $P+1$ partitions where the extra one partition is the partition that contains the solution (marked vertex).  \\

The contribution from this work is as follows. By applying the systematic dimensionality reduction technique via Lanczos algorithm, we extend 
the applicable graphs from complete graphs, complete bipartite graphs and star graphs \cite{novo2015systematic} to uniform complete multi-partite graphs. 
We extend a reduction scheme to transform an arbitrary $N$ by $N$ adjacency matrix $H_a$ of a uniform complete multi-partite graph into a 3 by 3 reduced Hamiltonian 
that has fast transport between its two lowest eigenenergy states. We further parameterize the coupling factor based on the configuration of a given  
uniform complete multi-partite graph to keep the CTQW search optimal. \\

The remainder of the article is organized as the following. In section \ref{sect:recap}, we first summarize
the notion of invariant subspace discussed in \cite{novo2015systematic}.  In section
\ref{sect:GeneralP}, we apply the method to analyze optimality of uniform complete P-partite graphs. 
In section \ref{sect:HamilSearchG} we further develop theorems to show (a) how to choose the correct 
coupling factor based on the given parameters (configuration) on a reduced graph and (b) the optimality is 
preserved once transformed back to the original graph.  By adding additional constraints to our finding, 
we recover many useful examples such as complete graphs, star graphs and complete bipartite 
graphs in section \ref{sect:Examples}. The reduced Hamiltonian is slightly different for each of these three cases because 
there are transitions among partitions that behave differently for each case.
Finally in section \ref{sect:discussion}, we draw our conclusion.

\section{Invariant Subspace of a Quantum Walk}\label{sect:recap}	%
Continuous-time quantum walk on a graph is a quantum dynamical process governed by a tight binding Hamiltonian.
Given a graph $G(V,E)$ (characterized by the vertex set V and the edge set E), one constructs the corresponding
CTQW model by first defining a Hilbert space with state $\kets{i}$ from node $i$ in $V$.  In most cases and in this study, 
the tight binding Hamiltonian is defined as
\begin{eqnarray}
\bras{i} H_a \kets{j} = \left\{ \begin{array}{cc}
1, \,\, (i,j) \in E \\
0. \,\, \text{otherwise}
\end{array}\right.
\end{eqnarray}
Alternatively, $H_a$ is simply called the adjacency matrix of the unweighted graph.

A time-evolved wave function on the graph is given by
\begin{eqnarray}
\kets{\psi(t)} & = & \exp(-i H_a t) \kets{\psi(0)} \nonumber \\
& = & \sum_{n=0}^{\infty} \frac{(it)^n}{n!} H_a^n \kets{\psi(0)} \nonumber \\
& = & \sum_{n=0}^{\infty} \frac{(it)^n}{n!} \kets{\psi^{(n)}(0)}.
\end{eqnarray}
Due to the finite dimensionality of the Hilbert space, the number of independent states 
$\mathcal{I}(H_a, \kets{\psi(0)}) \equiv span\{\kets{\psi^{(n)}(0)}=H^{n}_a\kets{\psi(0)}\}$ 
generated from the unitary dynamics (equivalent to repeated actions of the Hamiltonian) is 
bounded by $\vert V \vert$, the cardinality of vertex set.  
Following Ref.~\cite{novo2015systematic}, we designate $\mathcal{I}(H_a, \kets{\psi(0)})$ as the 
invariant subspace with respect to $\kets{\psi(0)}$.
When the Hamiltonian features certain symmetries, the invariant subspace could be much smaller than $\vert V \vert$.
Let $\mathcal P$ be the projection onto $\mathcal{I}(H_a, \kets{\psi(0)})$, one finds the same unitary dynamics can 
be generated by an effective Hamiltonian $\mathcal P H_a \mathcal P = H_{ra}$, i.e $\exp(-i H_a t) \kets{\psi(0)} = \exp(-i H_{ra} t) \kets{\psi(0)}$ 
for all time $t$.  In the following sections, we should apply this concept to identify the invariant subspace of a marked element $\kets{w}$ in multi-partite 
graphs and study the properties of CTQW in the reduced Hilbert space with an effective Hamiltonian $H_{ra}$.

\section{Search in Uniform Complete Multi-Partite Graphs}\label{sect:GeneralP}%
In this section, we first describe the procedures it requires to perform the dimentionality redution and CTQW construction 
based on the reduced dimension and the chosen coupling factor $\gamma$. We then further show that CTQW based on the chosen coupling factor $\gamma$ will 
still preserve its quadratic speed-up, i.e. remaining optimal. The reduction and coupling factor $\gamma$ determination process is as the following. 
\begin{algorithm}
\caption{Mechanism: Dimensionality Reduction and Coupling Factor Determination  \label{alg:MechG2}}
\begin{algorithmic}
\REQUIRE  A UCPG G of arbitrary size with one marked element $\kets{\omega}$\\
\ENSURE $\kets{\omega}$ can be found efficiently by CTQW.
\STATE {\bf Start of process}
\STATE $\bullet$  Dimensionality Reduction: Construct the reduced 3 by 3 Hamiltonian $H_{ra}$ by use of Lanczos algorithm on the $N$ by $N$ adjacency matrix $H_a$ based on a UCPG G
\STATE $\bullet$  Hamiltonian Construction: Construct CTQW Hamiltonian $H_{seek} = -\gamma H_{ra} - \kets{\omega}\bras{\omega}$.
\STATE $\bullet$  Basis Change: Express $H_{seek} = H^{(0)}+ H^{(1)}$ in the eigenbasis $(\kets{\omega}, \kets{e_1}, \kets{e_2})$ of $H^{(0)}$ by applying perturbation theory
\STATE $\bullet$  CTQW Initialization: Determine coupling factor $\gamma$ to induce fast transport between two lowest eigen energy states $\kets{\omega}$ and $\kets{e_1}$ in $H_{seek}$
\STATE $\bullet$  Existence of Constant Overlap: Demonstrate the initial starting state $\kets{s}$ and $ \kets{e_1}$ have a non-exponentially small overlap such that $\kets{s}$ can move
$\kets{\omega}$ efficiently via $\kets{e_1}$ and the optimality (quadratic speed-up) is preserved.
\STATE {\bf End of process}
\end{algorithmic}
\end{algorithm}
\subsection{Dimensionality Reduction}
\begin{figure}[th]
\centering
\includegraphics[scale=.3]{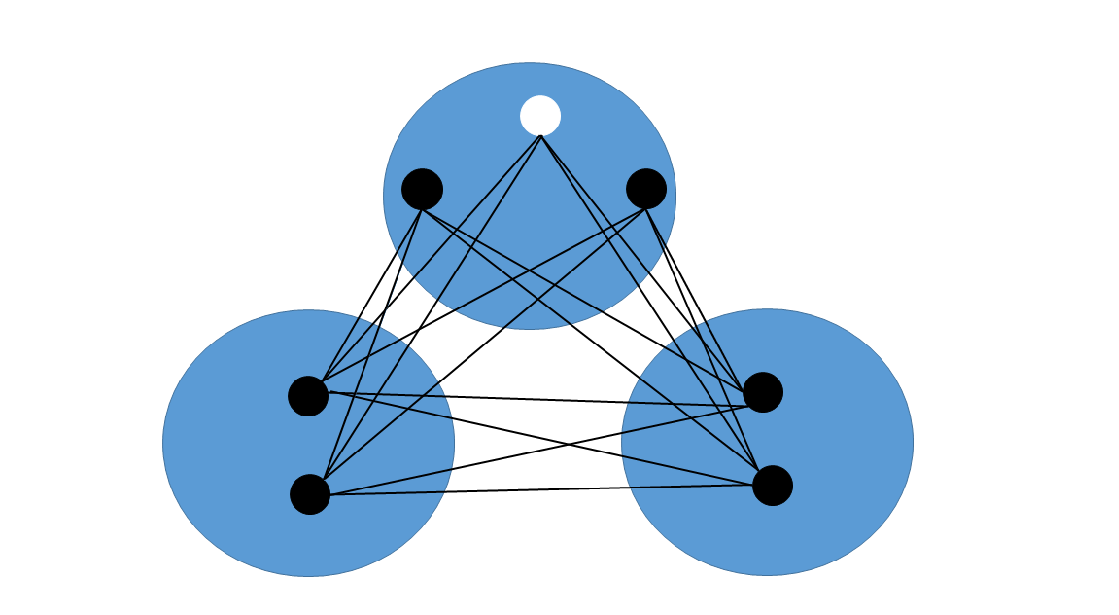}
\caption{A UCPG graph $G(V_0, V_1, V_2)$ where $m_0 = 3$ and $m_1 = 2$. The white element is the marked element $\kets{\omega}$ that resides in partition $V_0$.} 
\label{fig:UCPG}
\end{figure}

A uniform complete P-paritite graph (UCPG) can be denoted as $G(V_0, V_1, ..., V_P)$. 
It is a graph with $P+1$ partitions of vertices with the following properties: (1) each vertex $v_i$ in vertex partition $V_j$ connects to all other vertices in vertex partition $V_k$ as long as $j \neq k$ (2) except vertex partition $V_0$, each of the vertex partitions has the same size. Let the size of the vertex partition $V_j$ be $m_j$, i.e. $m_j=|V_j|$. Then we know that for UCPG graph with $N$ vertices, it automatically satisfies that $ P\times m_1 + m_0 = N$ as $m_1 = m_2 =\cdots = m_P$. An example of UCPG is given at Fig. \ref{fig:UCPG} as below. 

Without loss of generality, let us assume the marked vertex $\kets{\omega}$ is in $V_0$ and $m_0 \gg 1$. Define the subspace that is spanned by $\kets{\omega}, \kets{S_{V_0 - \omega}}, \kets{S_{{V}_1}}, \cdots,  \kets{S_{{V}_P}}$. With renormalization, we have 
\begin{equation}
\kets{S_{V_0 -\omega}}  =  \frac{1}{\sqrt{m_0 -1}} \sum_{i \in V_0, i \neq \omega}\kets{i}, \quad 
\kets{S_{{V}_i}}  =  \frac{1}{\sqrt{m_i}} \sum_{j \in S_{{V}_i}, i \neq 0}\kets{j}
\end{equation}
The adjacency matrix Hamiltonian $H_{a}$ of a given UCPG graph can thus be written in the basis states 
$(\kets{\omega}, \kets{S_{V_0 - \omega}}, \kets{S_{{V}_1}}, \cdots,  \kets{S_{{V}_P}})$ and it
behaves as the following:
\begin{eqnarray}
H_{a}\kets{\omega} & = & \sum_{j = 1}^P \sqrt{m_j}\kets{S_{{V}_j}}, \quad \quad
H_{a}\kets{S_{V_0 -\omega}}  = \sum_{j = 1}^P \sqrt{(m_0-1) m_j}\kets{S_{{V}_j}}\\
H_{a}\kets{S_{{V}_i, i \neq 0}} & = &	\sqrt{m_i}\kets{\omega}+ \sqrt{(m_0-1)(m_i)}\kets{S_{V_0 -\omega}} + \sum_{j, j\neq i, j \neq 0}^P \sqrt{m_i m_j}\kets{S_{{V}_j}}.
\end{eqnarray}

By use of Lanczos algorithm and the fact that partitions not containing $\kets{\omega}$ have the same size, the reduced adjacency 
Hamiltonian $H_{ra}$  in the $(\kets{\omega}, \kets{S_{V_0 - \omega}}, \kets{S_{\bar{V}_0}})$ basis is
\begin{equation} \label{eqn:Hra}
H_{ra} =
\begin{bmatrix}
0 & 0 & \sqrt{N-m_0} \\
0 & 0 & \sqrt{(N-m_0)(m_0-1)} \\
\sqrt{N-m_0} & \sqrt{(N-m_0)(m_0-1)} &  N-m_0-m_1\\
\end{bmatrix}
\end{equation}

\noindent
where $\kets{S_{\bar{V}_0}}  =  \frac{1}{\sqrt{N-m_0}}\sum_{j =1}^{P}\sqrt{m_j}\kets{S_{{V}_j}}$.

\subsection{Hamiltonian Construction and Basis Change}\label{sect:HamilConBasis}
For simplicity, let us define $\alpha = \frac{m_0}{N}$ and $\alpha_1 = \frac{m_1}{N}$. Since $H_{ra}$ expressed in 
the $(\kets{\omega}, \kets{S_{V_0 - \omega}}, \kets{S_{\bar{V}_0}})$ basis captures the same dynamics as $H_a$, the Hamiltonian of 
a CTQW can be defined as \cite{childs2004spatial}

\begin{equation}\label{eqn:HCQW}
H_{seek} = -\gamma H_{ra} - \kets{\omega}\bras{\omega}
\end{equation} where $\gamma$ is the coupling parameter between connected vertices. By Eqn.(\ref{eqn:Hra}, \ref{eqn:HCQW}), 
we know $H_{seek} = H^{(0)}+ H^{(1)} $ in the $(\omega,  S_{V_0 - \omega} , S_{\bar{V}_0})$ basis is 
\footnote{Clear that $\alpha_1 = (1-\alpha)/P$} \footnote{Entry (2,3) at $H^{(0)}$ is thus $-\gamma N ((1-\alpha) - (1-\alpha)/P) = -\gamma (N-m_0 - m_1)$}
\begin{equation}\label{eqn:H1H2}
H^{(0)} =
\begin{bmatrix}
-1 & 0 & 0 \\
0 & 0 & -\gamma N \sqrt{\alpha (1-\alpha)} \\
0 & -\gamma N \sqrt{\alpha (1-\alpha)} &  -\gamma N((1-\alpha)-\frac{P\alpha_1^2}{1-\alpha})\\
\end{bmatrix}
\end{equation}

\begin{equation}
H^{(1)} =
\begin{bmatrix}
0 & 0 & -\gamma  \sqrt{(1-\alpha) N} \\
0 & 0 & 0 \\
-\gamma \sqrt{(1-\alpha) N}& 0 & 0\\
\end{bmatrix}. 
\end{equation}

Prior to proceeding further, it is worth noticing that the format of this reduced Hamiltonian differs from the format derived in \cite{novo2015systematic} for a complete bipartite graph. 
The difference is the existence of a self-loop entry for the basis vector $ S_{\bar{V}_0}$. It later propagates in $H_{seek}$ and $H^{(0)}$. Because of this entry, in order to 
do systematic dimensionality reduction, it imposes a stronger constraint of equal size for partitions that do not contain the solution. We address this issue in order to generalize the 
result shown in \cite{novo2015systematic} for UCPG. As verified in section \ref{sect:Examples},  we know our generalization does encompass the result from  \cite{novo2015systematic}. \\

In the remaining of the section, we introduce Theorem \ref{thm:newbasis}, Lemma \ref{lma:facts} and  Theorem \ref{thm:gammavalue}. 
The relationships among them provide the foundation for showing the optimality preserving of the underlying CTQW. The
optimality preserving is explained in subsection \ref{sect:HamilSearchG}.  Theorem \ref{thm:newbasis} provides us the technique to construct the reduced Hamiltonian $H_{seek}$ 
in the eigenbasis $(\kets{\omega}, \kets{e_1}, \kets{e_2})$ of $H^{(0)}$. Lemma \ref{lma:facts} discovers important properties of Hamiltonian $H_{seek}$ written in the 
eigenbasis $(\kets{\omega}, \kets{e_1}, \kets{e_2})$ to be used in Theorem \ref{thm:gammavalue}.  Theorem \ref{thm:gammavalue} shows the necessary condition 
for fast transport to occur in $H_{seek}$ by tuning the coupling factor $\gamma$. \\

Now we prove Theorem \ref{thm:newbasis} to show how to express a reduced Hamiltonian $H_{seek}$ in the basis of its major matrix via perturbation theory. 
For simplicity, let us simply call $H_{seek}$ as $H$ in the theorem. \\

\theorem \label{thm:newbasis}
Given a reduced Hamiltonian $H = H^{(0)}+ H^{(1)}$ in the $(\kets{\omega},  \kets{b_1}, \kets{b_2})$ basis where
\begin{center}
\begin{equation}\label{h1h2}
H^{(0)} =
\begin{bmatrix}
-1 & 0 & 0 \\
0 & 0 & v_1 \\
0 & v_1 &v_3\\
\end{bmatrix}
, \quad
H^{(1)} =
\begin{bmatrix}
0 & 0 & v_2 \\
0 & 0 & 0 \\
v_2& 0 & 0\\
\end{bmatrix}
\end{equation}
 \end{center}
$v_1$ and $v_2$ are negative numbers and $v_3$ is a non-positive number where $v_1/v_2 = \sqrt{N \alpha} \geq 1$. Let  
the eigenvectors basis of  $H^{(0)}$ be $(\kets{\omega}, \kets{e_1}, \kets{e_2})$. We choose $\kappa = \frac{v_3}{v_1}\geq 0$ and $\beta_\pm =\frac {\kappa \pm \sqrt{\kappa^2 + 4}}{2}$, 
then we know eigenvector 
$\kets{e_1} =  \frac{(\kets{b_1} + \beta_+ \kets{b_2})}{\sqrt{1 + \beta_+^2}}$  and eigenvector 
$\kets{e_2} =  \frac{(\kets{b_1} + \beta_- \kets{b_2})}{\sqrt{1 + \beta_-^2}}$ where the corresponding eigenvalues are  $\lambda_\pm = v_1 \beta_\pm$. 
$H$ can thus be written in the $(\kets{\omega}, \kets{e_1}, \kets{e_2})$ eigenbasis as 

\begin{equation}\label{HSKe1e2}
H =
\begin{bmatrix}
-1 & v_2 \frac{\beta_+}{\sqrt{\beta_+^2 +1}} & v_2 \frac{\beta_-}{\sqrt{\beta_-^2 +1}} \\
\frac{v_2 \sqrt{\beta_+^2 + 1 }}{\beta_+ - \beta_-} & \lambda_+ & 0 \\
\frac{-v_2 \sqrt{\beta_-^2 + 1 }}{\beta_+ - \beta_-} & 0 & \lambda_-\\
\end{bmatrix}. 
\end{equation}
\begin{proof}
It is clear to see that $\kets{e_1}$ and $\kets{e_2}$ are both vectors of linear combination of 
$\kets{b_1}$ and $\kets{b_2}$. Without loss of generality, let $\kets{e'} = \kets{b_1} +\beta \kets{b_2}$ be an eigenvector 
of $H^{(0)}$ with eigenvalue $\lambda$.
After some calculation we  obtain $\lambda = {\beta v_1}$ where
\begin{equation}\label{eqn:betavalue}
\beta = \frac {\kappa \pm \sqrt{\kappa^2 + 4}}{2}, \quad  \kappa = \frac{v_3}{v_1}.
\end{equation}

For simplicity, let $\beta_{+}$ be $\frac {\kappa + \sqrt{\kappa^2 + 4}}{2}$ and $\beta_{-}$ be $\frac {\kappa - \sqrt{\kappa^2 + 4}}{2}$. 
By renormalizing the eigenvectors $\kets{e'_1} = \kets{b_1} +\beta_+ \kets{b_2}$, $\kets{e'_2} = \kets{b_1} +\beta_- \kets{b_2}$, 
we have 
\begin{equation}{\label{eqn:H0e1e2}}
\kets{e_1} = \frac{\kets{e'_1}}{\sqrt{\beta_+^2 +1}}, \quad
\kets{e_2} = \frac{\kets{e'_2}}{\sqrt{\beta_-^2 +1}}
\end{equation}
such that 
\begin{equation}
H^{(0)}\kets{e_1} = \lambda_+ \kets{e_1}, \quad H^{(0)}\kets{e_2} = \lambda_- \kets{e_2}
\end{equation} where 
\begin{equation}\label{lambdapm}
\lambda_{\pm} = \beta_{\pm}v_1.
\end{equation}

In the $(\kets{\omega},  \kets{b_1}, \kets{b_2})$ eigenbasis, from Eqn.(\ref{h1h2}) we know $H^{(1)}\kets{b_1} = 0$,  
$H^{(1)}\kets{\omega} = v_2\kets{b_2}$ and  $H^{(1)}\kets{b_2} = v_2\kets{\omega}$. To express $H^{(1)}$ in
the $(\kets{\omega}, \kets{e_1}, \kets{e_2})$ 
eigenbasis, by simple basis change,  we obtain 

\begin{equation}
H^{(1)}\kets{e_1} = v_2(\beta_+/(\sqrt{\beta_+^2 +1}))\kets{\omega}, \quad
H^{(1)}\kets{e_2} = v_2(\beta_-/(\sqrt{\beta_-^2 +1}))\kets{\omega}
\end{equation}

\begin{equation}
H^{(1)}\kets{\omega} =v_2\kets{b_2} = \frac{v_2 \sqrt{\beta_+^2 +1}}{\beta_+ -\beta_-}\kets{e_1} + \frac{-v_2 \sqrt{\beta_-^2 +1}}{\beta_+ -\beta_-}\kets{e_2}.
\end{equation}
Hence, the Hamiltonian $H$ can be expressed as shown in Eqn. (\ref{HSKe1e2}). 
\end{proof}

\lemma \label{lma:facts}
Given a derived reduced Hamiltonian $H$ written in the $(\kets{\omega}, \kets{e_1}, \kets{e_2})$ basis as shown in 
Theorem \ref{thm:newbasis}, we then know that (a) Hamiltonian $H$ is symmetric and 
(b) $\beta_+ > 0 > \beta_-$ and $\lambda_+ <0,  \lambda_- > 0$. 

\begin{proof}
With the value of $\beta_\pm$ as shown in Theorem \ref{thm:newbasis}, we know that 
\begin{equation}\label{betapm}
\beta_+ \beta_- = -1 
\end{equation}
and it immediately leads to the observation that 
\begin{equation}
\beta_+ (\beta_+ - \beta_-) = \beta_+^2 +1, \quad \beta_-(\beta_+ - \beta_-) = -(1+ \beta_-^2).
\end{equation}
With this observation, we can immediately conclude that 
\begin{equation}
v_2 \frac{\beta_+}{\sqrt{\beta_+^2 +1}} = \frac{v_2 \sqrt{\beta_+^2 + 1 }}{\beta_+ - \beta_-}, \quad
v_2 \frac{\beta_-}{\sqrt{\beta_-^2 +1}} = \frac{-v_2 \sqrt{\beta_-^2 + 1 }}{\beta_+ - \beta_-}.
\end{equation} 
Therefore, the property (a) that $H$ is symmetric is proved. For property (b), since $\sqrt{\kappa^2 + 4} > \kappa > 0 $, we immediately have 
$\beta_+ > 0 > \beta_-$. And with the fact that $v_1< 0$ and $\lambda_{\pm} = \beta_{\pm}v_1$, we can also immediately conclude that  $\lambda_+<0$ and $\lambda_- >0$ . 
\end{proof}

\normalfont
\noindent
For simplicity, let 
\begin{equation}\label{delta1delta2}
\delta_1 = v_2 \frac{\beta_+}{\sqrt{\beta_+^2 +1}},  \quad
\delta_2 = v_2 \frac{\beta_-}{\sqrt{\beta_-^2 +1}}.
\end{equation} By use of Lemma \ref{lma:facts},
$H$ can be written in the $(\kets{\omega}, \kets{e_1}, \kets{e_2})$ basis as \\
\begin{equation}\label{HSKe1e2delta}
H =
\begin{bmatrix}
-1 & \delta_1 & \delta_2 \\
 \delta_1& \lambda_+ & 0 \\
 \delta_2 & 0 & \lambda_-\\
\end{bmatrix}
\end{equation}
where $\kets{\omega}$ and $\kets{e_1}$ can form the basis for the two states of the lowest eigenvalue. \\

\theorem \label{thm:gammavalue}  Given a Hamiltonian $H$ in the form shown in Lemma \ref{lma:facts}, 
it is desirable to have $\lambda_+ = -1$ such that $\kets{\omega}$ and $\kets{e_1}$ form the basis for the 
two states of the lowest eigenvalue. Since $v_1= -\gamma N(\sqrt{\alpha (1- \alpha)})$ then the degeneracy between site energies of $\kets{\omega}$ and $\kets{e_1}$ 
facilitates transport between these two low energy states, 
hence $\gamma = (N\sqrt{\alpha (1- \alpha)}\beta_+)^{-1}$. The transport between $\kets{\omega}$ and $\kets{e_2}$ is 
prohibited since $\delta_2$ is much smaller than $\lambda_-$.

\begin{proof} Since we desire to have faster transport between the lowest eigen energy states, we need to set 
\begin{equation}\label{eqn:lambdaplus}
\lambda_+ = v_1\beta_+ = -1.
\end{equation}
With the fact that
$v_1 = -\gamma N(\sqrt{\alpha (1- \alpha)})$, we need to set 
\begin{equation}\label{eqn:gamma}
\gamma = (N\sqrt{\alpha (1- \alpha)}\beta_+)^{-1} 
\end{equation}
From Eqn.(\ref{delta1delta2}, \ref{HSKe1e2delta}) and $\lambda_-$ in Eqn.(\ref{lambdapm}), we know $\delta_2$ is 
much smaller than $\lambda_-$ because
\begin{equation}\label{eqn:delta2small}
\frac{\delta_2}{\lambda_-} = \frac{v_2}{v_1 \sqrt{(\beta_-^2 +1)}}=\frac{1}{\sqrt{\alpha N (\beta_-^2 +1)}}< \frac{1}{\sqrt{\alpha N}}\footnote{since $m_0 \gg 1$, $\alpha = m_0/N$, then $\alpha N \gg 1$ }. 
\end{equation}
\end{proof}

\normalfont
For a given UCPG G, by use of Eqn.(\ref{eqn:H1H2}-\ref{h1h2}), 
we can properly bound $\kappa$ as 
\begin{equation}\label{eqn:kappabound}
\kappa =\frac{v_3}{v_1} = \frac{(1-\alpha) - \frac{P \alpha_1^2}{1-\alpha}}{\sqrt{\alpha(1-\alpha)}}= \frac{\sqrt{(1-\alpha)}(1- \frac{1}{P})}{\sqrt{\alpha}}, \quad  0 \leq \kappa < \sqrt{\frac{1-\alpha}{\alpha}}.
\end{equation}

\subsection{From Existence of Constant Overlap to Optimality Preserving}\label{sect:HamilSearchG}
For a search space of size $N$, classical search has the complexity of $O(N)$. Quantum walk search provides a quadratic speed-up $O(\sqrt{N})$ in comparison to its 
classical counterpart. Please note that the complexity is for the number of calls to a single step of a search operation. For instance, in Grover it is the number of Oracle calls. 
In the remainder of this subsection, we will show that the quadratic speed-up (optimality) remains with the $\gamma$ chosen based on Theorem \ref{thm:gammavalue}. \\

\begin{figure}[th]
\centering
\includegraphics[scale=.5]{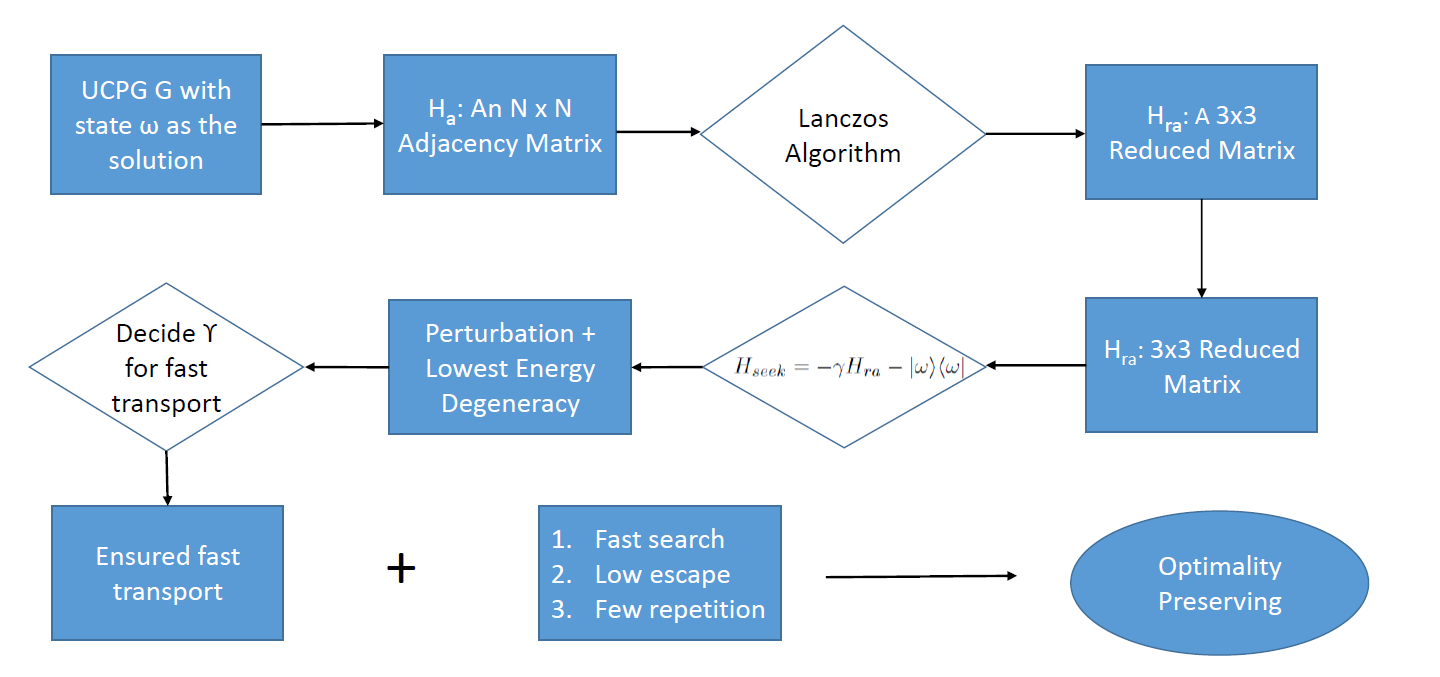}
\caption{The procedure from systematic dimensionality reduction, basis change, fast transport and finally optimality preservation.} 
\label{fig:flow}
\end{figure}

For a given UCPG G, the processing flow described in algorithm \ref{alg:MechG2} can be shown as a flow chart in Fig.\ref{fig:flow}.

By using the the theorems and lemma from subsection \ref{sect:HamilConBasis}, $H_{seek}$ can be expressed as Eqn.(\ref{HSKe1e2}) in the the 
eigenbasis $(\kets{\omega}, \kets{e_1}, \kets{e_2})$ of $H^{(0)}$. By rewriting Eqn.(\ref{eqn:H0e1e2}) using applying Eqn.(\ref{eqn:Hra}, \ref{eqn:HCQW}) and 
Theorem \ref{thm:newbasis}, we know
\begin{equation}
\kets{e_1} =  \frac{(S_{V_0 - \omega} + \beta_+ S_{\bar{V}_0})}{\sqrt{1 + \beta_+^2}}, \quad 
\kets{e_2} =  \frac{(S_{V_0 - \omega} + \beta_- S_{\bar{V}_0}))}{\sqrt{1 + \beta_-^2}}
\end{equation}   
where $\beta_\pm =\frac {\kappa \pm \sqrt{{\kappa} ^2 + 4}}{2}$. \\

For a CTQW based on $H_{seek}$, we need to decide the value of coupling 
parameter $\gamma$ to ensure the optimal performance of the underlying quantum walk is preserved. If the coupling parameter $\gamma$ is wrongly
chosen, the underlying CTQW search might not remain optimal, i.e. its quadratic speed-up might 
be lost. The determination process of correct $\gamma$ is shown in Theorem \ref{thm:gammavalue}. 
Theorem \ref{thm:loweigenG} is an extension of Theorem \ref{thm:gammavalue} to various cases with respect to the values of variable $P$ and variable $\alpha$. 

\theorem \label{thm:loweigenG} 
Given a UCPG $G=(V_0, V_1, \cdots, V_P)$ and its adjacency matrix Hamiltonian $H_a$ in the $(\kets{\omega},  \kets{b_1}, \kets{b_2})$ basis where $N =\sum_{i=0}^P |V_i|$, we can obtain the reduced search Hamiltonian $H_{seek}$ in a new eigenbasis $(\kets{\omega}, \kets{e_1}, \kets{e_2})$ by use of Theorem \ref{thm:newbasis} for constructing the underlying CTQW. We can then use Thereom \ref{thm:gammavalue} to determine the coupling factor $\gamma = (N\sqrt{\alpha (1- \alpha)}\beta_+)^{-1}$. The chosen $\gamma$ ensures the underlying CTQW remains optimal. \\ 

\begin{proof} 
There are two aspects that we need to address to show that the optimality $O(\sqrt{N})$ 
is preserved. One (1) is fast search speed and low escaping speed while the other one (2) is the overlap
between $\kets{e_1}$ and the initial system state $\kets{s}$ (a uniform superposition) as it determines how many times we need to 
repeat the experiment. \\

The search speed is determined by the dynamics between fast transport non-solution $\kets{e_1}$ and solution state $\kets{\omega}$, i.e. $\kets{e_1} \rightarrow \kets{\omega}$. The degenerate eigenspace formed by $\kets{\omega}$ and $\kets{e_1}$ captures the dynamics between those two states.
The escape speed is from solution $\kets{\omega}$ to undesirable non-solution states $\kets{e_2}$. \\

From Eqn.(\ref{HSKe1e2delta}), 
we know that $\delta_1$ is responsible for the search speed and $\delta_2$ is responsible for escape speed. In Eqn.(\ref{eqn:delta2small})We have shown that $\delta_2$ is small with respect to $\lambda_{-}$, the escape speed is small. By use of Eqn.(\ref{delta1delta2}), we know that 
\begin{equation}
|\delta_1| = \bras{e_1} H_{seek}\kets{\omega} = |\frac{v_2 \beta_+}{\sqrt{\beta_+^2 +1}}| = |\frac{-1}{\sqrt{\alpha N (\beta_+^2 +1)}}|
\end{equation}
because $v_2 = -\gamma \sqrt{N(1-\alpha)}$ and $\gamma = (N\sqrt{\alpha (1- \alpha)}\beta_+)^{-1}$. Hence, we obtain the running time 
\begin{equation}\label{eqn:runtime}
T_{run} = \pi\sqrt{\frac{\alpha N (\beta_+^2 +1)}{2}}.
\end{equation}

Let us verify that the running time $T_{run}$ remains optimal in different settings of UCPG G when the 
coupling factor $\gamma$ is chosen based on Theorem \ref{thm:gammavalue}. Briefly speaking, with a fixed search space of size $N$, 
the configuration of a UCPG G is controlled by variable P and variable $\alpha$. We will discuss different settings based on those two variables. \\

\noindent
Case 1: $ P = 1$ \\
This is a typical complete bipartite graph as seen in \cite{novo2015systematic}. We immediately know that $\kappa = 0$ since $\alpha_1 = 1- \alpha$ from Eqn.(\ref{eqn:kappabound}). 
This leads to $\beta_+ = 1$ from Eqn.(\ref{eqn:betavalue}). Because of that, no matter what value of $\alpha$ is, $T_{run}$ at Eqn.(\ref{eqn:runtime}) holds its quadratic speed-up. \\

\noindent
Case 2: $2 \leq P \leq N-1$ and $\alpha \propto \frac{1}{N}$\\
By Eqn.(\ref{eqn:kappabound}), we know that $\kappa \propto \sqrt{N-1}$ and by  Eqn.(\ref{eqn:betavalue}), we know $\beta_+ \propto  \sqrt{N-1}$. By plugging in the values 
of $\alpha$ and $\beta_+$, $T_{run}$ at Eqn.(\ref{eqn:runtime}) still holds its quadratic speed-up. \\

\noindent
Case 3: $2 \leq P \leq N-1$ and $\alpha \propto 1$ (such as $\frac{N-1}{N}$)\\
By Eqn.(\ref{eqn:kappabound}), we know that $\kappa \propto 1/\sqrt{N-1}$; by  Eqn.(\ref{eqn:betavalue}), we know $\beta_+ \propto  \frac{(1/\sqrt{N-1})+\sqrt{(1/N-1)+4}}{2} \simeq 1$ 
when $N$ is large. By plugging in the values of $\alpha$ and $\beta_+$, $T_{run}$ at Eqn.(\ref{eqn:runtime}) still holds its quadratic speed-up. \\

\noindent
Case 4:  $2 \leq P \leq N-1$ and  $\alpha$ is some constant (non-extreme values): \\
Immediately we know $\kappa$ and $\beta_+$ are some constants that would not affect the complexity.  Hence, $T_{run}$ at Eqn.(\ref{eqn:runtime}) still holds its quadratic speed-up.\\ 

\noindent
However, the $T_{run}$ above assumes that we start the search from eigenstate $\kets{e_1}$ to find $\kets{\omega}$, i.e. $\kets{e_1} \rightarrow \kets{\omega}$, which is not the case because we 
start from $\kets{s}$. Hence, at $T_{run}$ the success probability of observing $\kets{\omega}$ is the overlap between $\kets{e_1}$ and $\kets{s}$. The success probability is \footnote{Simply compute their inner product and we know that
$\kets{s} = \frac{\kets{\omega} + \sqrt{m_0 -1}\kets{S_{V_0 - \omega}} + \sqrt{N-m_0}\kets{S_{\bar{V}_0}}}{\sqrt{N}}$}
\begin{equation}\label{enq:successProb}
P_{O} = |\braket{e_1}{s}|^2  = \Big| \frac{\sqrt{\frac{\alpha}{\beta_+^2} - \frac{1}{\beta_+^2 N}} + \sqrt{1-\alpha}}{\sqrt{1 + \frac{1}{\beta_+^2}}}\Big|^2. 
\end{equation}
Therefore $1/P_{O}$ is the number of times we need to repeat the experiment. We need to show that $P_{O}$ is some constant such that it would not affect the total complexity under the Big $O$ notation. 
By examining the four cases listed earlier and putting the values of $\alpha$ and $\beta_+$ into Eqn.(\ref{enq:successProb}),  we know that $P_O$ remains as some constant that is not exponentially small. \\

Since the total runtime is 
\begin{equation}
T_{run} \times \frac{1}{P_{O}}
\end{equation}
where $T_{run}$ holds quadratic speed-up and $\frac{1}{P_{O}}$ is some constant that is not large (not scaling with $N$), the complexity still holds the quadratic speed-up. Therefore, we know that the chosen $\gamma =  (N\sqrt{\alpha (1- \alpha)}\beta_+)^{-1}$ ensures the underlying CTQW remains optimal.  
\end{proof}
\normalfont

\section{Specific Examples} \label{sect:Examples} %
Given a uniform complete P-partite graph (UCPG) $G = (V_0, V_1, \cdots, V_P)$, we know  $P*m_1 + m_0 = N$ and 
the $H_{ra}$ for an UCPG can be simplified as 
\begin{equation} \label{eqn:HraUPPG}
H_{ra} =
\begin{bmatrix}
0 & 0 & \sqrt{N-m_0} \\
0 & 0 & \sqrt{(N-m_0)(m_0-1)} \\
\sqrt{N-m_0} & \sqrt{(N-m_0)(m_0-1)} & (N-m_0)(1 - \frac{1}{P})\\
\end{bmatrix}
\end{equation}
since $ (N-m_0-m_1) = (N-m_0)(1 - \frac{1}{P})$, 
then we can obtain Eqn.(\ref{eqn:HraUPPG}) from Eqn.(\ref{eqn:Hra}). 

In this section we translate the UCPG into the three extreme graphs; complete graph, bipartite graph and star graph, 
as demonstrated in \cite{novo2015systematic} by 
simply choosing the right value for $P$ and $m_0$. By showing the equivalence of the adjacency matrix in each case, 
we generalize those three cases with our UCPG interpretation. 
Since the adjacency matrices are equivalent, what follows is that we will have the same search Hamiltonian 
and the tuning factor $\gamma$ discussed in subsection \ref{sect:HamilConBasis} and \ref{sect:HamilSearchG}.  

\begin{itemize}
\item Complete Graph: In this case we have $m_0 = m_1 = 1$, and $ P = N - 1$. 
By applying to Eqn.(\ref{eqn:HraUPPG}), we obtain the adjacency matrix 
in the $(\kets{\omega}, \kets{S_{V_0 - \omega}}, \kets{S_{\bar{V}_0}}) $ basis  
$$
H_{ra} =
\begin{bmatrix}
0 & 0 & \sqrt{N-1} \\
0 & 0 & 0 \\
\sqrt{N-1} & 0 & (N-2)\\
\end{bmatrix}
$$
which is exactly the reduced Hamiltonian of a complete graph case with $N$ nodes  \cite{novo2015systematic}. 

\item Complete Bipartite Graph: In this case, we have $P=1$ since we only have two partitions, one contains the 
marked vertex and one does not. By applying to Eqn.(\ref{eqn:HraUPPG}), 
we obtain the adjacency matrix in the $(\kets{\omega}, \kets{S_{V_0 - \omega}}, \kets{S_{\bar{V}_0}}) $ basis  
$$
H_{ra} =
\begin{bmatrix}
0 & 0 & \sqrt{m_1} \\
0 & 0 & \sqrt{(m_1)(m_0-1)} \\
\sqrt{m_1} & \sqrt{(m_1)(m_0-1)} & 0\\
\end{bmatrix}
$$\\ 
which is exactly the bipartite case \cite{novo2015systematic} with $N$ nodes cut into two partitions while 
one partition is of size $m_0$ and the other is of size $m_1$ where $m_0 + m_1 = N$. 

\item Star Graph: In this case, the center is the partition $V_1$ that does not contain the marked vertex while the 
satellite states, including the marked vertex, are in partition $V_0$. In this case, we have $m_0 = N-1$ and $m_1 = 1$, and $P = 1$. 
By applying to Eqn.(\ref{eqn:HraUPPG}), we obtain the adjacency matrix  in the $(\kets{\omega}, \kets{S_{V_0 - \omega}}, \kets{S_{\bar{V}_0}}) $ basis 
$$
H_{ra} =
\begin{bmatrix}
0 & 0 & 1 \\
0 & 0 &  \sqrt{N} \\
1 & \sqrt{N} & 0\\
\end{bmatrix}
$$ where one partition that contains the marked vertex is the set of $N-1$ satellite nodes while the other partition 
is the center node of size $1$. We use $\sqrt{N}$, instead $\sqrt{N-1}$, when $N$ is large. This 
is exactly the reduced Hamiltonian of a star graph case with $N$ nodes \cite{novo2015systematic}.  
\end{itemize}

\section{Discussion} \label{sect:discussion} %
The notion of invariant subspaces\cite{novo2015systematic} of continuous-time quantum walk (CTQW) problems is a 
powerful technique that simplifies the analyses of various quantum walk related studies such as the spatial search algorithm, 
quantum transport, and quantum state transfer. In essence, it maps a spatial search algorithm to a transport problem on a 
reduced graph. The dimensional reduction is purposely constructed to preserve the dynamical
evolution of a walker.  Hence, any quantum walker optimization on a reduced graph guarantees
an optimization on the original graph.  In this work, we apply this technique  to deduce 
an appropriate coupling factor for the underlying CTQW to run optimally (to keep the quadratic speed-up with 
running time $O(\sqrt{N}$)) for a spatial search.  We generalize the result in \cite{novo2015systematic} from 
complete graphs (CG), complete bipartite-graphs (CBG)
and star graphs (SG) to uniform complete P-partite graphs (UCPG). It is clear that UCPG could be non-regular or regular 
based on the constraints we impose. More specifically,  we (1) derive the formula for the coupling factor $\gamma$ and 
(2) show that CTQW constructed based on our choice of coupling factor will remain optimal. \\

To demonstrate the validity of our main results in section \ref{sect:GeneralP}, we show how to translate a UCPG graph to CG, CBG and SG.
It is clear we have the hierarchy $CG, CBG, SG \subset UCPG$ by simply adding constraints on the general class to form more limited classes as explained in section \ref{sect:Examples}. 
We further verify our work by examining our conclusion on UCPG with the CG, CBG and SG cases shown in \cite{novo2015systematic}. 
Our coupling factor formula acting on the reduced UCPG echoes the results shown in the CG, CBG and SG cases. \\

The proof of the optimality is two-fold. The speed of the CTQW is based on (1) the transport efficiency 
between the two lowest energy eigenstates (one is the marked state $\kets{\omega}$ and the other 
state is $\kets{e_1}$) and (2) the overlap between the initial state $\kets{s}$ and the $\kets{e_1}$
in the invariant subspace. We showed that the transport efficiency preserved the quadratic speed-up 
and the overlap is some constant that does not scale with the inverse $N$ (that is to say
it would not be exponentially small). Therefore, the CTQW based on a coupling factor determined 
by our formula will remain optimal. \\

\section{Acknowledgments}
		C.~C. gratefully acknowledges support from the State University of New York Polytechnic Institute. 

\normalfont
\bibliographystyle{plain}
\bibliography{Bibtex}
	 \end{document}